\documentclass{aa}

\usepackage[english]{babel}

\usepackage{amsmath}
\usepackage{graphicx}
\usepackage[colorlinks=true, allcolors=blue]{hyperref}
\usepackage{xspace}

\title{Diffraction in the ASPIICS coronagraph: observations and modeling}
\author{S.~Shestov\inst{\ref{ROB},\ref{CSL}}
\and
A.~N.~Zhukov\inst{\ref{ROB},\ref{MSU}}
\and
R.~Rougeot\inst{\ref{ESTEC}}
\and
C.~Aime\inst{\ref{UNice}}
\and
B.~Bourgoignie\inst{\ref{ROB}}
\and
L.~Dolla\inst{\ref{ROB}}
\and
N.~Britavskiy\inst{\ref{ROB}}
\and
S.~Fineschi\inst{\ref{INAF}}
\and
S.~Gun\'ar\inst{\ref{ASU_CAS}}
\and
P.~Lamy\inst{\ref{LATMOS}}
\and
M.~Mierla\inst{\ref{ROB},\ref{IGAR}}
\and
H.~Peter\inst{\ref{MPS}}
\and
P.~Rudawy\inst{\ref{UWr}}
\and
K.~Tsinganos\inst{\ref{UoA}}
}

  \institute{
  Solar--Terrestrial Centre of Excellence --- SIDC, Royal Observatory of Belgium, 1180 Brussels, Belgium\\
  \email{s.shestov@oma.be}
  \label{ROB}
  \and 
  Centre Spatial de Li\`ege, Universit\'e de Li\`ege, Av. du Pr\'e-Aily B29, 4031 Angleur, Belgium
  \label{CSL}
   \and
  Skobeltsyn Institute of Nuclear Physics, Moscow State University, 119991 Moscow, Russia
  \label{MSU}
\and
  European Space Research and Technology Centre, European Space Agency, Noordwijk, Netherlands
  \label{ESTEC}
  \and
  Universit\'e C\^{o}te d’Azur, Centre National de la Recherche Scientifique, Observatoire de la C\^{o}te d’Azur, UMR7293 Lagrange, Parc Valrose, 06108, Nice, France
  \label{UNice}
  \and
  National Institute for Astrophysics, Astrophysical Observatory of Torino, Pino Torinese, Torino, Italy
  \label{INAF}
  \and
  Astronomical Institute of the Czech Academy of Sciences, 251 65 Ond\v{r}ejov, Czech Republic
  \label{ASU_CAS}
  \and
  Laboratoire Atmosph\`eres et Observations Spatiales, 11 Boulevard d’Alembert, 78280 Guyancourt, France
  \label{LATMOS}
   \and
  Institute of Geodynamics of the Romanian Academy, 020032 Bucharest-37, Romania
  \label{IGAR}
 \and
  Max Planck Institute for Solar System Research, Justus-von-Liebig-Weg 3, 37077 G\"ottingen, Germany
  \label{MPS}
  \and
  Astronomical Institute, University of Wroc\l{}aw, Kopernika 11, 51-622 Wroc\l{}aw, Poland
  \label{UWr}
  \and
  University of Athens, Panepistimiopolis, 157 84 Zografos Athens, Greece
  \label{UoA}
   }

\begin{document}
\newcommand{\HRIEUV}{HRI\ensuremath{_\mathrm{EUV}}\xspace}
\newcommand{\cmC}{cm$^{-3}$\xspace}
\newcommand{\DEMTm}{\mathrm{DEM(T)}\ }
\newcommand{\Aime}{\citep[][]{2013A&A...558A.138A}\xspace}
\newcommand{\RR}{\citep{2017A&A...599A...2R}\xspace}
\newcommand{\Shest}{\citep{Shestov2018}\xspace}
\newcommand{\SZ}{SZ18\xspace}

\abstract
{ASPIICS is a giant-baseline visible light solar coronagraph, which relies on the millimetric positioning performance of the precision formation flying Proba-3 mission of the European Space Agency. Proba-3 was launched on 5 Dec 2024, and since then ASPIICS observes the solar corona with the field of view $(1.1-3)R_\sun$.  }
{Diffraction, in particular diffraction of solar disk light on the external occulter, is known to provide a major source of straylight in coronagraphs. We aim to analyze diffracted light visible in ASPIICS images, compare it with the analytical-numerical diffraction model reported earlier, and fine-tune the model. }
{We compare diffraction effects visible in ASPIICS data with simulated diffraction images; in particular, we  compare the geometrical properties and the radiometric signal. The properties of the diffraction described in previous works suggest how to fine-tune the model in order to achieve a better correspondence with the observations. }
{Early ASPIICS observations, where diffraction is pronounced, fully confirm all the qualitative properties of diffracted light suggested by the model. After fine-tuning of the model we see quantitative correspondence of the level of 30\% -- 50\%, depending on the configuration. }
{The performed analysis allows (a) to validate our analytical-numerical model and justify the assumptions, and (b) to estimate the contribution of the diffracted light in the  ASPIICS images. In the majority of the field of view the diffracted light is two orders of magnitude below the coronal signal. }

\maketitle

\nolinenumbers

\section{Introduction}
Diffraction plays an important role for solar visible light coronagraphy, where the solar corona is observed next to the extremely bright -- by at least 6 orders of magnitude brighter -- solar disk. Even in the case of externally occulted coronagraphs, the bright emission of the photosphere propagates to the region of the geometrical shadow behind the external occulter (EO), enters the optical system and creates straylight in the coronal images. Due to the relatively dominant brightness of the solar disk, even this small amount of diffracted light can be brighter than the corona itself. Thus, other diaphragms, such as an internal occulter (IO) and a Lyot stop, are used to reduce the diffracted light. 
The straylight is usually subtracted during on-ground data processing by using empirical approaches, such as the one based on a monthly averaged signal. At the same time, there are continuous efforts to calculate the diffracted light with more advanced models \citep{Rougeot2018, Aime2020}, or, improve our understanding of the background physics \citep{Wang:21,2025ApJ...982...58D}.

The amount of diffracted light increases with the reduction of EO overoccultation \citep{1978A&A....63..243F, Lenskii1988}. The ASPIICS coronagraph \citep[Association of Spacecraft for Polarimetric and Imaging Investigation of the Corona of the Sun, see][]{Zhukov2025} aboard the Proba-3 mission of the European Space Agency (ESA) utilizes the precise formation flying of the two mission spacecraft. The front spacecraft carries an external occulter with $R_\textrm{EO}=710$~mm at an approximate inter-satellite distance (ISD) of $\sim 144$~m in front of the aperture of the telescope mounted on the second spacecraft. The EO forms a shadow with a   geometrical umbra of $\sim 70$~mm in diameter, and the entrance aperture of the ASPIICS telescope is placed in the center with a high precision (below 1~mm), thus forming a giant externally-occulted coronagraph. This unique geometrical configuration results in the minimal field of view (FOV) of $\sim 1.1 R_\sun$ with the aim of achieving in the same time a sufficiently low amount of diffracted light which enters the telescope.

The amount of the diffracted light reaching the primary objective depends on several factors, such as the shape of the EO, the distance between the EO and the objective and the overoccultation angle  defining how much larger an occulter is compared to the solar disk that it is intended to block \citep{Lenskii1988, Bout2000}. In the beginning of the ASPIICS project a series of experimental works \citep{LAndini2010,Landini2011} have been carried out to identify the best shape of the occulter. Later, a series  of theoretical analyses \citep{Aime2013,RR17,Shestov2018} considered the intensity of the diffracted light entering the telescope, and calculated the two-dimensional pattern of the diffraction in the detector plane.

The Proba-3 mission has been successfully launched on 5 Dec 2024. Since then, the commissioning phase has been completed and the nominal mission phase is ongoing since July 2025 \citep{Zhukov2025wind}. The first observational data show that ASPIICS works well, observing the solar corona with an unprecedented combination of FOV, high spatial and temporal resolutions, and seemingly a low level of the diffracted and stray light. 
During calibration campaigns with telescope off-pointing, some diffracted light can be identified in the images. In this paper, we rely on the available images and compare the diffraction model and the observations. In Sect.~\ref{sec:layout} we present the optical configuration of ASPIICS, in Sect.~\ref{sec:model} we recall the details of the diffraction models and their main properties, and in Sect.~\ref{sec:images} we show early ASPIICS images and perform a preliminary comparison with the diffraction model. In Sect.~\ref{sec:refinement} we refine the model to better match the observations, in Sect.~\ref{sec:others} we analyze the influence of several factors, in Sect.~\ref{sec:regular} we analyze the level of the diffracted light in typical ASPIICS image, and finally in Sect.~\ref{sec:conclusions} we discuss the results and draw some conclusions.

\section{Optical and functional layout}
\label{sec:layout}
\begin{figure*}[!ht]
  \centering
  \includegraphics[width=16cm]{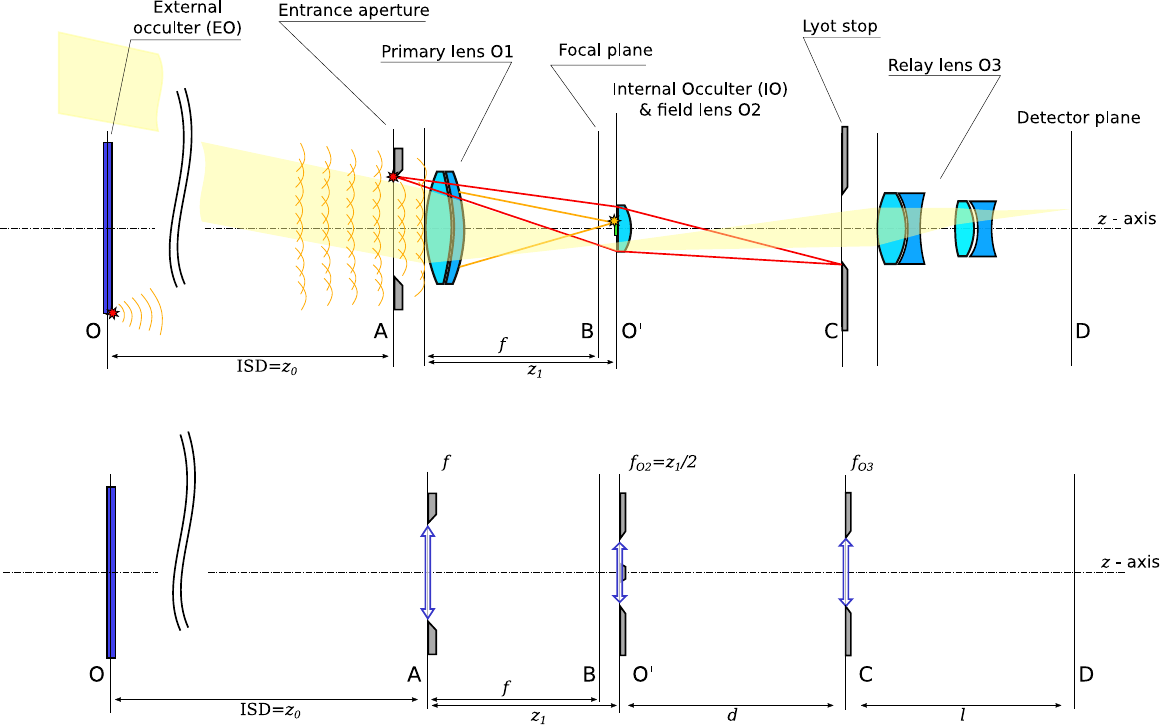}
  \caption{\textit{Top} panel: optical layout of the ASPIICS coronagraph. \textit{Bottom} panel: a simplified version of the optical design used in the diffraction model. See text for details. }
  \label{fig:layout}
\end{figure*}

The optical layout of ASPIICS has been reported in \citet{2015SPIE.9604E..0BG} and \citet{Galano2019}, while the functional layout for the diffraction modeling is extensively discussed in \citet{RR17} and \citet[][hereafter SZ18]{Shestov2018}. We provide a brief description here, with the optical and schematic layouts given in Fig.~\ref{fig:layout}.

The EO is situated in front of the telescope at a nominal distance $z_0\approx144.5$~m (or ISD); the exact distance can be configured, as it is adjusted during the year to compensate for solar angular size changes. The entrance pupil with $\varnothing 50$~mm is situated 120 mm in front of the primary objective (PO) with the design focal length $f_\mathrm{PO}=330.35$~mm. The field lens O2 is situated slightly behind the primary focus, at a distance $\Delta\approx 0.76$~mm. A non-transparent annular mask -- the internal occulter (IO) -- with external radius $R_\mathrm{IO}=1.752$~mm is deposited on the flat front surface of O2. A group of relay lenses O3 projects the image from the primary focus to the detector, providing an effective focal length $f=734.6$~mm. The Lyot stop is situated in front of the O3 lens, having an effective open diameter which is 97\% of the entrance aperture image. A $2048 \times 2048$~pixels CMOS-based detector with the pixel size 10~$\mu$m provides a pixel plate scale of $\sim 2.817$~arcsec/pixel, and the full FOV of $1.6\degr$ along the side of the detector. Another feature that distinguishes ASPIICS from other coronagraphs is relatively narrow annular vignetting zone of $\sim 25$~pixel wide, such that above $0.31\degr$ vignetting disappears. Thus taking into account central symmetry and vignetting by the IO, the FOV ranges from $0.293\degr$ to $0.8\degr$, and commonly referred to as $1.1-3.0 R_\sun$ (with an implied assumption $R_\sun=16$~arcmin).

The diffraction model uses a slightly simplified layout, with the entrance aperture and the PO being on the same plane A, the primary focus is on the plane B, IO and O2 are on the plane O${^\prime}$, the Lyot stop and O3 on the plane C, and finally the detector is on the plane D. The models use a Fourier-optics formalism, and thus a simplification of infinitely-thin lenses is used. Another adopted simplification is that the image is projected from the B plane to the D plane with a unit magnification, and the aperture is projected to the C plane with also a unit magnification. This sets simple relations between distances and focal lengths. In the diffraction model the distance between the O3/Lyot stop and O2/IO is denoted by $d=z_1$, thus fixing $f_{O2}=z_1/2$; and, the distance between the C and D planes is denoted by $l$.    

\section{Main properties of the diffraction suggested by the models}
\label{sec:model}
The three diffraction models -- \citet{Aime2013}, \citet{RR17}, \SZ\ -- sequentially complement each other. In \citet{Aime2013} the diffraction was analyzed in the plane of the entrance aperture of ASPIICS,  after propagating behind the EO. In \citet{RR17} the model was extended to include the rest of the optical system, and the diffracted light was calculated in the detector plane. Both studies assumed a symmetrical configuration, where the Sun-EO line coincided with the optical axis of the telescope. \SZ extended these models by introducing misalignments, in particular studying the cases when the entrance aperture of the telescope was not centered within the umbra-penumbra pattern, and when the optical axis of the telescope was not parallel to the Sun-EO line. They pointed out a rather significant influence of the misalignments on the intensity and distribution of the diffracted light.

The essential features of the diffraction models are as follows:
\begin{enumerate}
    \item The Sun is considered as a spatially extended source, a spatial sampling is established which includes center-to-limb darkening.
    \item For each plane-parallel wave coming from a sampled element of the Sun, diffraction on a knife-edge EO is considered, and the two-dimensional complex amplitude of the diffracted wavefront is calculated in the plane A.
    \item A Fourier-optics formalism is used to propagate the wavefront through ASPIICS, from the entrance aperture to the detector.
    \item The squared amplitudes of the wavefronts in the detector plane are summed for all the sampled elements of the Sun to give the diffracted light pattern.
\end{enumerate}
For step 2 the diffracted wavefront is calculated as follows: the co-axial wave is expressed analytically using a Hankel integral for Huygens-Fresnel diffraction on a circular occulter (see Eq.~\ref{PsiA00} in Appendix~\ref{sec:Ap_method}). The integral is numerically calculated for every pixel of the aperture plane. For the off-axis waves two additional transformations are applied: a  geometrical translation and an introduction of an additional phase tilt (Eq.~\ref{Tab}) for the whole wavefront. 

From the numerical analyses of the model, the following properties of the diffracted light have been identified:
\begin{enumerate}
    \item In the focal plane B of the primary objective the diffraction pattern has a shape of a bright ring with exponentially decreasing wings. The bright ring has a single-peak shape and a radius that corresponds to the radius of the EO, geometrically projected onto the plane B. The ring is broadened due to being out of focus \citep[see red curve in Fig.~6 in][]{RR17}.
    \item In the detector plane D the diffraction has a shape of a bright ring with two-peaks core with exponentially decreasing wings \citep{RR17}.
    \item In the plane D the bright ring remains always in the same geometrical location that corresponds to the projection of the IO to the detector; the actual positions of the Sun and the EO do not influence its location. The position of the diffraction ring coincides with the vignetting zone -- the region where the coronal signal gradually drops due to obscuration by the IO (\SZ).
    \item The core of the bright ring has two peaks, separated by $\sim 0.05~R_\sun$ (see Fig.~11 in \citealt{RR17}, also Sect.~4.1 in \citealt{2022A&A...665A.109T}).
    \item The radius of the core of the ring and its intensity depend on $R_\mathrm{IO}$: the larger the $R_\mathrm{IO}$, the lower the intensity, and the radius of the bright ring is proportional to $R_\mathrm{IO}$.
    \item Introduction of misalignments makes the bright ring non-uniform in intensity along the polar angle $\varphi$ -- it becomes brighter at one side, and dimmer at the opposite side (see Fig.~10 in \SZ). The effect depends on the size of the IO and the amplitude of the misalignment; the change of intensity can be as large as two orders of magnitude at 25~arcsec misalignments.
    \item The single-peaked diffracted ring can be observed in the detector plane D, if the IO does not block the bright diffracted ring (e.g. if the telescope is significantly off-pointed with respect to the EO).
\end{enumerate}

\SZ have shown that two major types of misalignments are possible: (a) when the telescope aperture is not co-centered with the umbra-penumbra pattern (panels \textit{b}, \textit{c}, and \textit{d} of their Fig.~5), and (b) when the telescope is pointing not to the center of the EO, but slightly aside (while the aperture can still be perfectly co-centered with the umbra-penumbra pattern; panel \textit{e} of their Fig.~5). In \SZ the first misalignment is called shift of the Sun, and the second one is called tilt of the telescope. Apparently, any physical configuration, such as the relative position and the orientation of the Sun, the EO and the telescope, can be expressed as a combination of the telescope tilts (in two directions) and solar shifts (also in two directions). 

\section{ASPIICS images and observations of diffraction}
\label{sec:images}
After the ASPIICS data are transferred to the Science Operation Center (SOC) at the Royal Observatory of Belgium, the data are  decompressed, the images are assembled and various calibrations are applied. Depending on the number of calibration steps, the images are called Level-1, Level-2, and Level-3. They are all stored as FITS files. Level-1 images are the raw images with all the metadata such as position of the satellites on the orbit, pointing information of the telescope, temperatures stored in the headers. These images have units of DN (digital numbers). In Level-2 images basic effects, such as the detector dark current and bias, flatfield, and the detector nonlinearity are corrected. The images are normalized by the exposure time and radiometrically calibrated, thus acquiring units of mean solar brightness (MSB). However, the images are not Sun-centered or de-rotated, thus orientation of the solar north can be arbitrary. Finally, in Level-3 images Sun-centering and de-rotation are performed, such that the solar north is up, and also the files registered with different exposure times $t_\mathrm{exp}$ are merged into a single high-dynamic range image \citep{Zhukov2025wind,Zhukov2025}.

\citet{Shestov2021} compared intensities of the expected coronal signal and various straylight components using provisional characteristics of the telescope and available models for the diffraction and other types of straylight -- ghost light and scattered light \citep{2019A&A...622A.101S}. They concluded that the diffracted light was the main contributor of the straylight. Furthermore, it was shown that under nominal conditions (misalignments below 25~arcsec) the intensity of the diffracted light should be $\sim$ 2 orders of magnitude below the coronal signal. \citet{2019A&A...626A...1R} and \citet{2022A&A...665A.109T} also confirmed small contribution of the diffraction.
While the diffracted light should be present in the images, thus slightly increasing the measured intensity, separating the diffraction from the coronal signal might be challenging.

There are two possibilities to directly observe the diffracted light in ASPIICS images:
\begin{itemize}
  \item under large off-pointings of order 0.5--1$^\circ$, when the bright diffracted ring is not blocked by the IO, but is still visible in the FOV. In this case the image of the diffracted ring is re-projected from plane B to plane D;
  \item under moderate off-pointings of order of 20--40~arcsec, when the two-peaks diffracted ring becomes visible in the vignetting zone.
\end{itemize}

Such observations were performed during Proba-3 orbits\footnote{Coronal observations by Proba-3 last up to 5.5 hours around the orbit apogee, which can extend over two calendar days. It is convenient to identify the coronagraphy windows by the number of the orbit.} 152 on 8 April 2025 (moderate off-pointing), and 352 on 19 September 2025 (large off-pointing).

\subsection{Large off-pointing $\sim 0.8^\circ$ during orbit 352 and diffraction from the external occulter}
\label{sec:large_offpointings}
\begin{figure}[!ht]
  \centering
  \includegraphics[width=9cm]{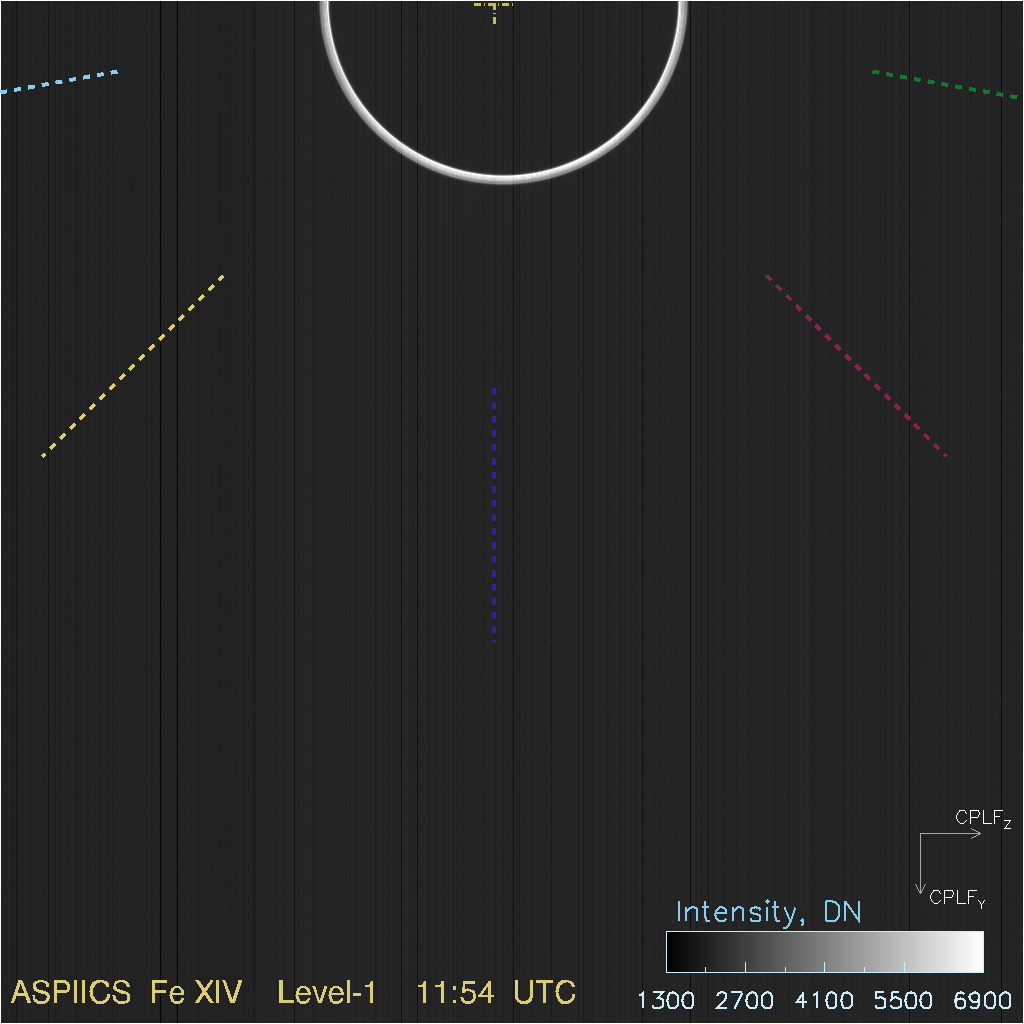}
  \caption{ASPIICS image in Fe XIV filter with $\sim 0.8^\circ$ off-pointing registered during  orbit 352. The image is Level-1, that is neither any calibration steps have been applied, nor $t_\mathrm{exp}$-normalized or radiometrically calibrated; it has units of DN. The diffracted light that came to the instrument is not blocked by the IO, thus the bright diffraction ring forming at the primary focus B is re-projected to the detector. The color dashed lines show the directions along which radial profiles have been taken (see Figure~\ref{fig:orbit352_profiles}).}
  \label{fig:orbit352_raw}
\end{figure}

During orbit 352 the telescope was off-pointed by $\sim 0.8^\circ$, while the aperture of the telescope was kept co-centered with the umbra-penumbra pattern. Thus, this off-pointing corresponds to the tilt of the telescope as considered in \SZ. The images were registered using the narrow-band Fe~XIV filter with a range of small exposure times from $10^{-4}$ to $1$~s, to not overexpose. An example of an image registered at 11:54~UTC is given in Fig.~\ref{fig:orbit352_raw}. This is a Level-1 image without any calibrations applied. The bright ring is produced by the diffraction from the EO, which is partially outside the FOV. The solar center (yellow cross) coincides with the EO center with a precision better than few pixels (see below). The vertical lines with slightly varying intensity, seen across the whole image, are due to the bias of the detector. This bias is almost completely removed in the Level-2 images. The color dashed lines denote the position of the radial profiles, discussed and analyzed below. In the bottom part of the figure we give the color scale and show the orientation of the main reference frame of the telescope -- CPLF (Coronagraph PayLoad Frame). Rotation of the telescope around CPLF$_y$ axis is called pitch, rotation of the telescope around CPLF$_z$ is called yaw. Thus, the off-pointing in this image corresponds to yaw, since the telescope was rotated around CPLF$_z$, and the objects in the image shift along CPLF$_y$. 

Proba-3 is equipped with a set of systems, that actively control relative position of the satellites, to ensure the aperture of the telescope is well-centered with the umbra-penumbra pattern. One of the subsystems, responsible for the measurements is the Shadow Position Sensor (SPS). According to SPS the relative displacement of the telescope aperture is of order of 100~$\mu$m. Should there be any discrepancy between the directions to the center of the EO and the Sun, it would immediately imply de-centering of the aperture with respect to the umbra-penumbra pattern: 5 arcsec (that is $\sim 2$-pixels) discrepancy corresponds to $\Delta \approx 144 \times \tan(5\arcsec) = 3.5$~mm displacement, that would also introduce unsymmetry to the diffracted light picture. 

\begin{figure}[!ht]
  \centering
  \includegraphics[width=9cm]{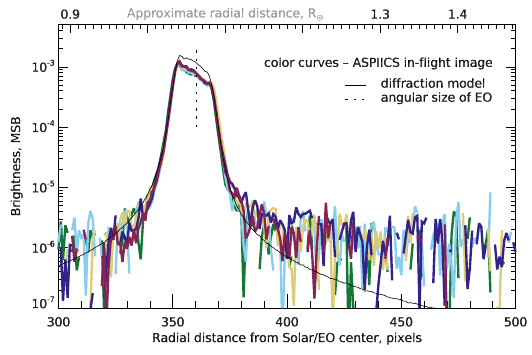}
  \caption{Comparison of radial profiles of the diffraction extracted from the calibrated Level-2 ASPIICS image (color curves) and the diffraction model (black curve). The calibrated image was obtained from the one shown in Figure~\ref{fig:orbit352_raw}. The vertical dotted line corresponds to the angular size of the EO for orbit 352.}
  \label{fig:orbit352_profiles}
\end{figure}

We calculated diffraction for this configuration using the numerical model from \SZ with the tilt angle of $0.8^\circ$, and using the ISD=145.017~m and $R_\sun=15.92$~arcmin (actual values at the time of the image acquisition). Additionally, after analysing various solar limb darkening profiles (see Sect.~\ref{sec:limb_darkening}), we have chosen the one reported by \citet{1977SoPh...51...25P}. Comparison of the radial profiles of the calibrated Level-2 ASPIICS image and the diffraction model is given in Fig.~\ref{fig:orbit352_profiles}. The profiles demonstrate good correspondence between the model and the observations. The maximum brightness in the diffraction model is $\sim 30\%$ higher than the observed one, however the width of the diffracted ring and reduction of the signal by two orders of magnitude in just seven pixels is correctly reproduced. The profiles in the ASPIICS image become flat at heights $>400$~pixels ($>1.15R_\sun$) because of extremely low signal at such distances and the dominant contribution of noise. Correspondence of the profiles in different directions also confirms the assumption of the co-centering of the Sun and the EO.

\subsection{Moderate off-pointing during orbit 152 and the double-peaked fringe}
During orbit 152 ($\mathrm{ISD}=144.63$~m, $R_\sun=15.96$~arcmin) the coronagraph satellite performed various manoeuvres that included lateral displacements of the telescope by $\pm3$~mm and afterwards off-pointings of the telescope by $\pm20$~arcsec, keeping the aperture of the telescope in the center of the shadow. The first type of manoeuvres, which mainly corresponds to solar shift in terms of \SZ, did not reveal any significant changes in the images (there was a very slight displacement of the EO satellite in the image due to parallax). The second type of maneouvres corresponds to the tilt of the telescope in terms of \SZ, and was performed along two directions  -- pitch and yaw. In the images corresponding to these off-pointings the coronal structures were moving by $\pm 7$~pixels with respect to each other. 

\begin{figure}[!ht]
  \centering
  \includegraphics[width=9cm]{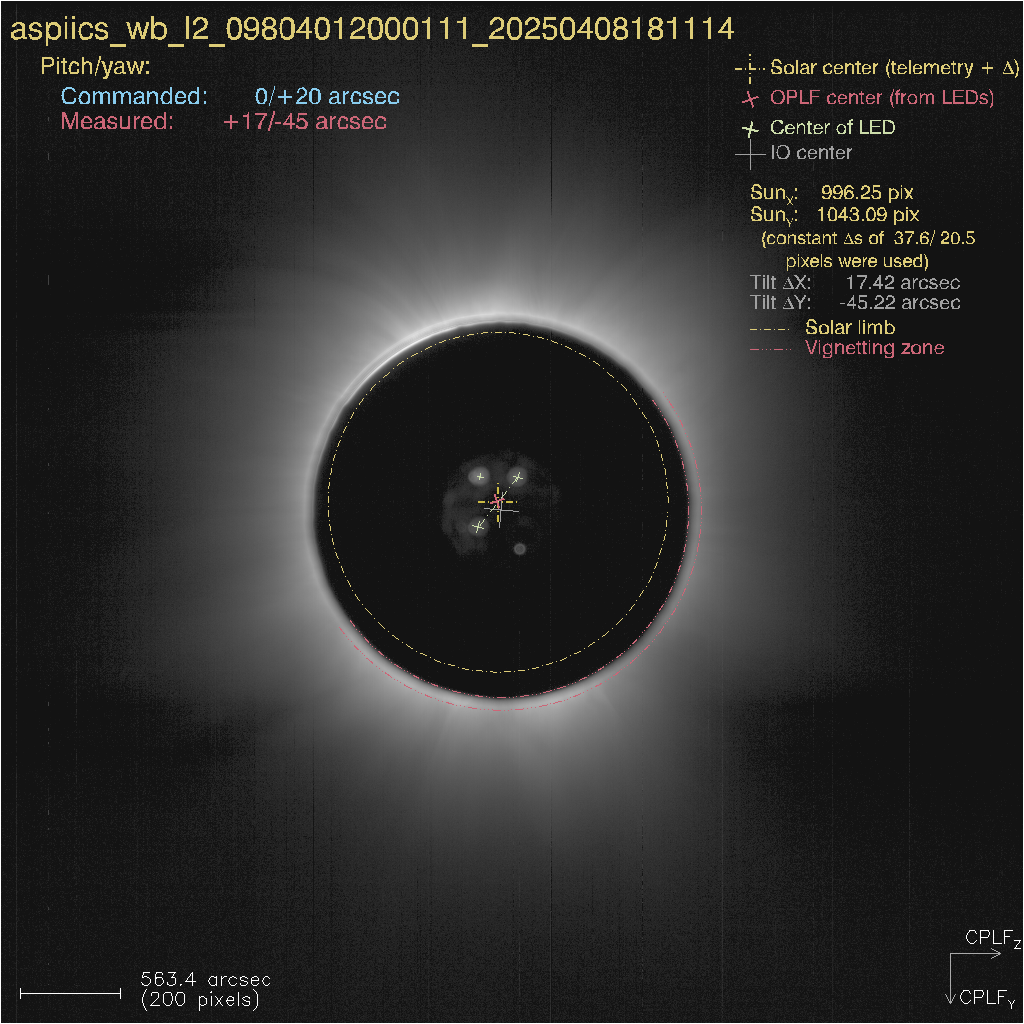}
  \caption{ASPIICS image in the wideband filter taken during orbit 152 at 18:11 UTC on 8 Apr 2025. The estimated off-pointing of ASPIICS is +17/-45~arcsec in pitch/yaw. See text for details. An animated gif with images with different off-pointings is available on-line. }
  \label{fig:orbit152_overview}
\end{figure}

We have found that both the nominal pointing -- with the commanded pitch=yaw=0 -- and with off-pointings had significant additional displacement with respect to the EO center, the same displacement for all the cases. Because of it the actual off-pointings amounted to 48~arcsec in some cases. An example Level-2 image, registered at 18:11 UTC is given in Fig.~\ref{fig:orbit152_overview}, when  commanded off-pointing was 20~arcsec in yaw. In order to calculate the actual off-pointing, we assume the solar center coincides with the center of the EO with good precision\footnote{When the manuscript was under review, this was verified using position of the stars in the FOV for coronal observations on 15 August 2025} (see explanation in Sect.~\ref{sec:large_offpointings}), and the actual off-pointing is calculated as the distance between the center of the Sun/EO and the IO, and in this case it amounts to +17 and -45~arcsec in pitch/yaw. We note that the same correction works for all the images from this series. The direction of solar north is approximately horizontal, pointing towards left\footnote{The coronagraph spacecraft was rolled by 180$^\circ$ with respect to its nominal orientation}. In the central part of the image we mark several important features: the center of the internal occulter with a grey cross, the assumed center of the Sun with a dark yellow cross, the solar limb is denoted by dash-dotted yellow circle, the vignetted zone is denoted by two red dash-dotted lines (the annular $\sim25$~pixel zone). The three small light-yellow crosses near the image center denote the position of the OPSE LEDs\footnote{OPSE -- Occulter Position Sensor Emitter; LED -- Light-Emitting Diode. See \citet{Zhukov2025}} on the occulter satellite, which are seen through the central opening in the IO. The center of the external occulter is situated in the middle of the hypotenuse of the right triangle formed by the LEDs. 

\begin{figure*}[!ht]
  \centering
  \includegraphics[width=0.86\textwidth]{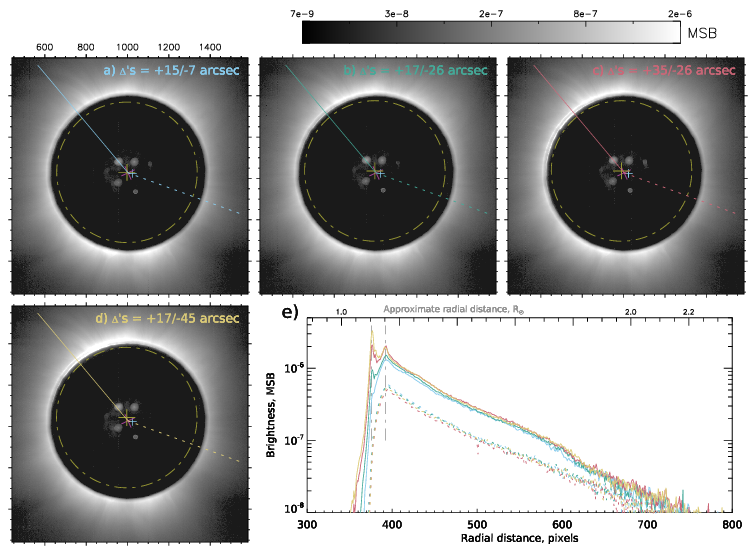}
  \caption{Images (panels \textit{a}--\textit{d}) and radial profiles (panel \textit{e}) taken during orbit 152 with various off-pointings. The estimated off-pointing in pitch and yaw are annotated in each panel. The solid and dashed lines highlight the directions along which the radial profiles have been measured. The crosses in each panel denote: blue cross -- center of the image, red cross -- center of the IO, yellow cross -- center of the Sun. Due to off-pointings the center of the Sun is different in each panel, while the center of the detector and the IO stay in the same place. The radial profiles (colors correspond to main images) were measured from the center of the IO. The log-like color scale is shown in the top.}
  \label{fig:orbit152_different files}
\end{figure*}

In Fig.~\ref{fig:orbit152_different files} we show the central part of several images with different off-pointings and radial profiles. In panels \textit{a}--\textit{d} we show the same part of the detector with the size $1144\times1144$~pixels, the estimated off-pointing is annotated in top of each panel. The two-peak diffraction is clearly visible in the upper-left part in the panels \textit{b}, \textit{c}, and \textit{d}. The solid and dashed lines in the images denote position of the profiles, which are shown in the panel \textit{e}. The colors in the plot correspond to the annotations in the images. The horizontal axis denotes the distance in pixels measured from the IO center. The two grey vertical dash-dotted lines with coordinates $\sim 376$ and $\sim 392$ pixels (angular separation $\sim 0.05 R_\sun$) denote approximate position of the diffraction peaks.

The position of the peaks of the double-peaked diffraction profile stays in the same place in different images. The intensity of the diffracted light is significantly larger in the direction where the solar limb and the edge of the EO come closer to the edge of the IO (that is along the solid lines), while in the opposite directions (along the dashed lines) the diffracted light has the intensity that is five times smaller. 

Inspection of the images with different commanded pitch and yaw reveals a good correspondence of the observed properties of the diffracted light to the diffraction model and its properties described in Sect.~\ref{sec:model}. While the coronal scene shifts across the image following the commanded off-pointings, the diffraction pattern stays in the same place. In the case of small off-pointings of order of $\sim 20$~arcsec (Fig.~\ref{fig:orbit152_different files}a), the diffraction is almost invisible in the images, however at larger off-pointings of order of $\sim 40$~arcsec (Fig.~\ref{fig:orbit152_different files}b, c, d), the double-peak structure starts to be pronounced in the top-left part close to the solar limb. 

\section{Refinement of the diffraction model using observations during orbit 152}
\label{sec:refinement}
In order to reproduce the observed properties of the diffracted light seen during orbit 152, we had to slightly modify the geometrical prescription of the diffraction model. In particular, we focused on the separation between the peaks of the double-peak profile of the diffracton ring, variation of the $r_\mathrm{IO}$\footnote{We denote as $r_\mathrm{IO}$ the IO radius expressed in pixels.} over the polar angle $\varphi$ and its mean value, and analyzed correspondence of the model to the images registered with different off-pointings.

\subsection{Separation between the two peaks of the diffraction ring}
\begin{figure*}[!ht]
  \sidecaption
  \centering
  \includegraphics[width=12cm]{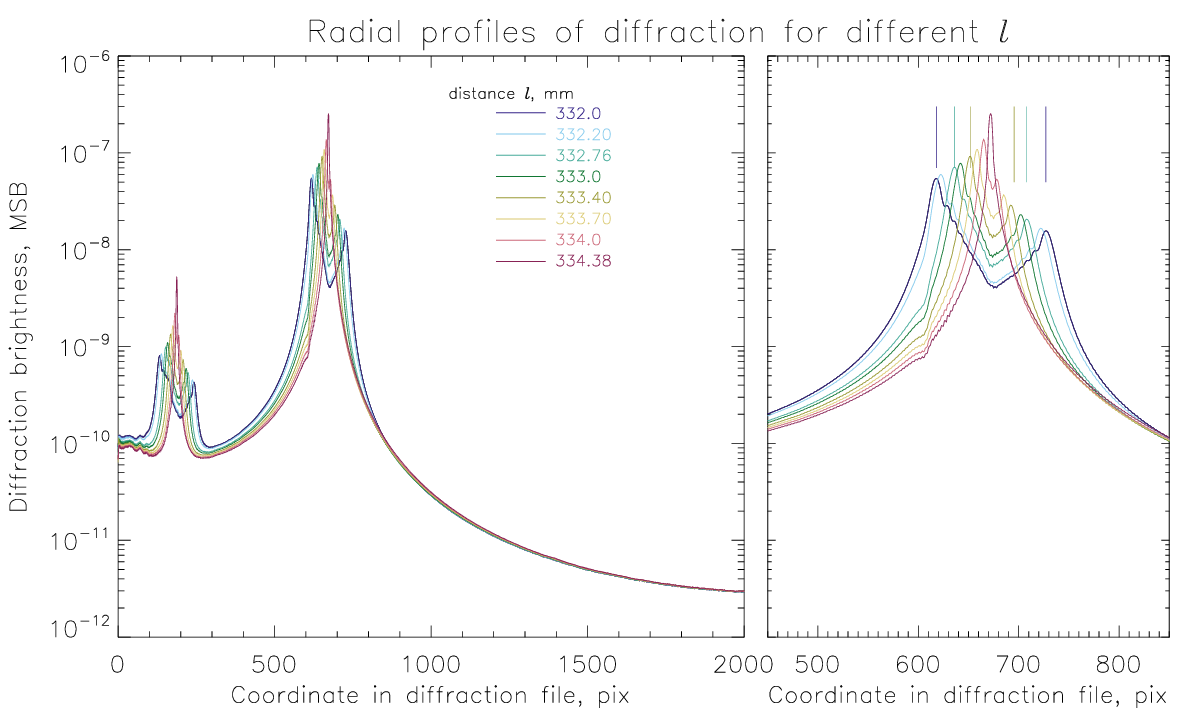}
  \caption{Radial profiles of diffraction for different distance $l$ used in the model. The full scale is given in the \textit{left} panel, while the zoomed region covering the radial distance 450--850 pixels is given in the \textit{right} panel.}
  \label{fig:dependence_on_l}
\end{figure*}

By analyzing the behaviour of the numerical model we have figured out that the separation between the diffraction peaks depended on the degree by which the IO is out-of-focus. The plane D is conjugated to the plane B as imaged by the O2 and O3 lenses, thus the image of the IO should be formed behind the D plane. We calculated the diffraction image for various distances $l$, varying the length in steps by a total amount of 2.4 mm. The radial profiles of the diffraction for various distances $l$ are given in Fig.~\ref{fig:dependence_on_l}, full scale on the left and zoomed region on the right. The vertical scale is expressed in MSB, while the horizontal scale is expressed in pixels of the original diffraction simulations. The vertical lines in the right panel approximately denote position of the peaks of individual simulations. For the minimal distance $l=332.0$~mm (violet curve), which was our original value used in \SZ, the separation between the peaks is larger than that in the observations. It decreases with the increase of $l$, and for $l=334.38$~mm (dark red curve) the two diffraction peaks merge into a single one, and the diffraction ring becomes almost perfectly focused. For $l=333.7$~mm the simulations were well representing the actually observed images. We note that apart from the changes in separation between the two peaks, other properties like the mean brightness, or the position of the center between the two peaks, do not change with the change of $l$. 

We have found that the radial position of the diffraction peaks was varying with the polar angle $\varphi$, with the variations of order of 1--2~pixels. Analysis of these positions for two files with the largest off-pointings showed, that position of both peaks $r_{1,2}(\varphi)$, as well as their average distance $\overline{r_\mathrm{1,2}}(\varphi)$, had the same functional dependence on $\varphi$. Moreover, $\overline{r_\mathrm{1,2}}(\varphi+180\degr)$ coincided with the measured radius of the internal occulter and its angular variation $r_\mathrm{IO}(\varphi)$, as derived from the on-ground calibration data. 

The presented analysis confirmed that the two diffraction peaks are equidistant from the average value, which corresponds to the radius of the IO; and the observed variation of $r_\mathrm{1,2}$ with $\varphi$ can be explained by the actual shape of the IO and imperfections of its shape. These variations, however, are of order of $\approx 0.5$\% of its radius; if translated to physical size they will be of order of $\delta R_\mathrm{IO}=9$~$\mu$m, which is extremely small. We have introduced the variable $r_\mathrm{IO}(\varphi)$ into the refined numerical model described in Section~\ref{sec:refined}.

\subsection{Conversion of scales}
The spatial and angular pixel scale in the original diffraction simulations depend on the physical and numerical sizes of the arrays used in the diffraction model \SZ. In order to be used with ASPIICS images, the spatial scale of the simulated diffraction files should be converted using some geometrical factors.

Such a conversion is straightforward for the plane B, since only one Fourier transform, representing the primary objective, is applied. For the plane B the the linear scale is calculated as $dx_\mathrm{B}=1/ (N \cdot (\mathrm{LA}/N)) \cdot \lambda f_\mathrm{PO}=f_\mathrm{PO} \lambda / \mathrm{LA} \approx 2.596$~$\mu$m/pixel, where $f_\mathrm{PO}$ is the focal length of the PO, $\lambda=0.550$~$\mu$m is the wavelength, $N=4096$ is the linear size of the array representing the A plane, LA=70~mm is assumed physical size of the array, and $\mathrm{LA}/{N}$ is spatial sampling of the A plane. For the plane D we need to take into account the effects of the O2 and O3 and the fact that we have adjusted $l$.

We  have also used another approach: we have introduced to the A plane several plane-parallel test waves tilted by angles 281.7, 563.4, 1408.6, 2253.8, and 2535.5 arcsec. These values correspond to the positions 100, 200, 500, 800, 900 pixels of  ASPIICS images with angular scale 2.8172 arcsec/pixel. Then we propagated the waves towards the detector with exactly the same formulas as the diffracted light, and measured position of their peaks in the simulated images. In these images the test waves appeared either as point-sources, or few-pixel-defocused sources (for the values of $l$ that significantly differed from the optimal value). We have figured out that the position of the test waves was always the same for different values of $l$; after re-scaling their exact locations coincided with the intended 100, 200, 500, 800, and 900 pixels if the factor ${\arctan{(dx_\mathrm{B}/ f_\mathrm{PO})}}/{2.8172}=0.5754 $ was applied. This reasoning justified the conversion of scales. 

After the spatial scale was established, we adjusted the physical size of the IO radius in the diffraction model to $R_\mathrm{IO}=1.742$~mm so that the position of the two diffraction peaks (which includes dependence on $l$ and on $r_\mathrm{IO}(\varphi)$) to match the position in the registered images after rescaling.

\subsection{Empirical coefficient for the intensity}
\label{sec:coeff}
We have found empirically that applying an additional multiplicative factor $k=1.2$ provides better results: after the diffraction was subtracted from the images the radial profiles had smaller peaks and dips in the vignetting zone. Necessity for this factor means that the diffraction model underestimates the  intensity of the diffracted light in the plane D. This contrasts with the results from Sect.~\ref{sec:large_offpointings}, however in that case the IO does not block the main diffraction ring. We discuss possible explanations in Sect.~\ref{sec:conclusions}.

\subsection{Refined model}
\label{sec:refined}
\begin{figure*}[!ht]
  \centering
  \includegraphics[width=16cm]{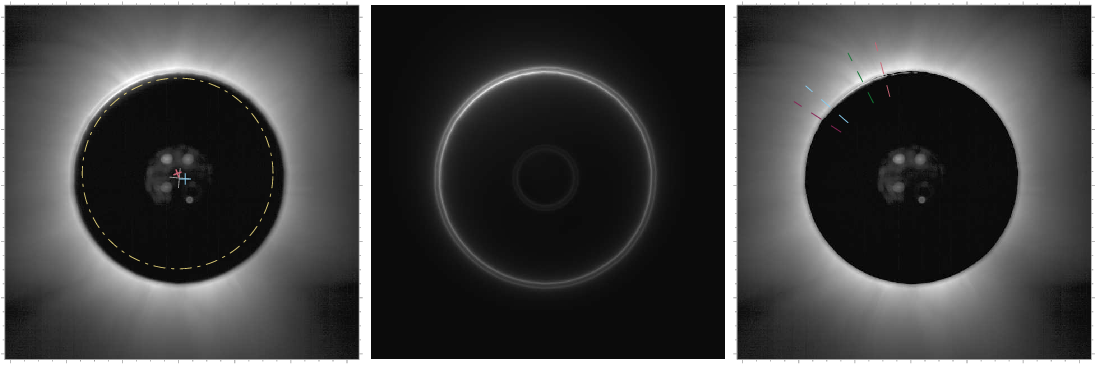}
  \caption{Comparison of the ASPIICS image (\textit{left}), the simulated diffraction image (\textit{middle}), and ASPIICS image after diffraction removal (\textit{right}) registered at 18:11 UTC (panel \textit{d} from Fig.~\ref{fig:orbit152_different files}). The simulated image was calculated for the tilt angles $45$~arcsec towards top and $17$~arcsec towards left (i.e. +17/-45~arcsec in pitch/yaw). In the right panel several direction are annotated, along which the radial scans are shown in Fig.~\ref{fig:sim_profiles}.}
  \label{fig:orig_vs_sim}
\end{figure*}

We have used the refined model to calculate diffraction for several ASPIICS images from orbit 152 (from the panels \textit{b}, \textit{c}, and \textit{d} in Fig.~\ref{fig:orbit152_different files}) for which the combinations of pitch/yaw angles were:  +17/-26 arcsec,  +35/-26 arcsec, and +17/-45 arcsec. In Fig.~\ref{fig:orig_vs_sim} we show the results for the image with +17/-45 arcsec off-pointing: the original image (\textit{left}), the simulated diffraction (\textit{middle}) and the image after diffraction is subtracted (\textit{right}). The central portion ($1264\times1264$ pixels) of the images is shown. 

\begin{figure*}[!ht]
  \centering
  \includegraphics[width=18cm]{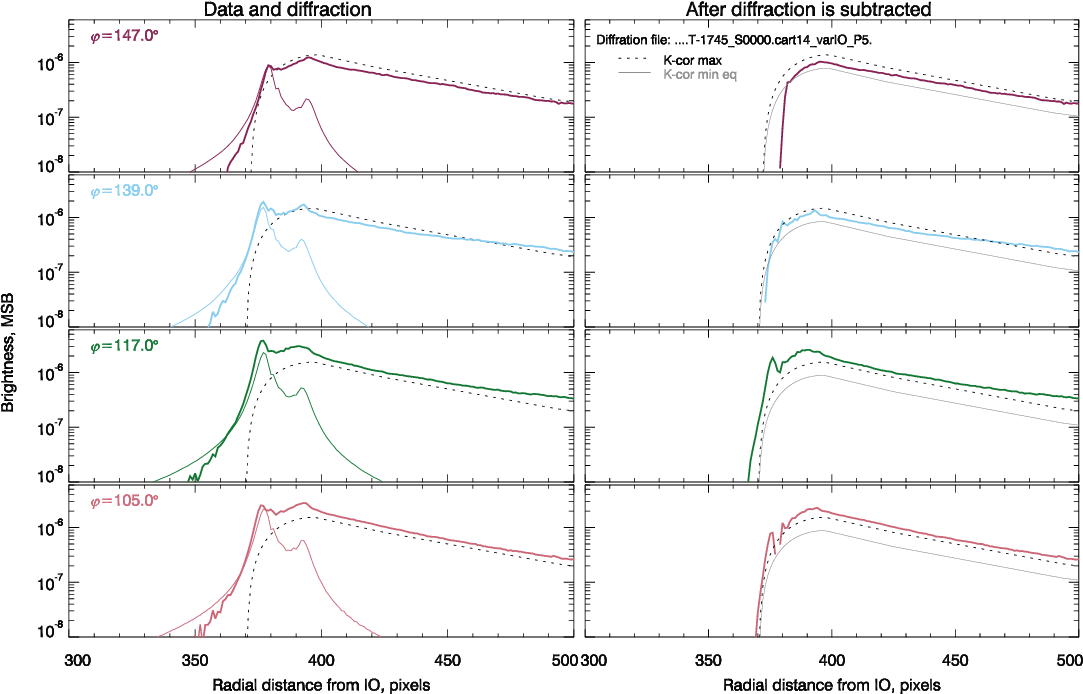}
  \caption{Radial profiles from the images presented in Fig.~\ref{fig:orig_vs_sim} with +17/-45~arcsec off-pointing. Different rows/colors correspond to different polar angles (shown in the right panel of Fig.~\ref{fig:orig_vs_sim}). \textit{Left} panels: radial profiles of the original ASPIICS image (thick color curves) and simulated diffraction (thin color curves); \textit{right} panels: radial profiles in the ASPIICS image after subtraction of diffraction (thick color curves). The horizontal axes are expressed in ASPIICS pixels and are measured from the IO center; the coordinates expressed in solar radii would be different in different panels, because solar center does not coincide with the IO. Gray lines correspond to K-corona during solar maximum (dashed line) and minimum (solid line) from \citet{Koutchmy2000}. These model curves are given as a reference to simplify comparison of different panels; in particular they correctly show drop of intensity due to vignetting. }
  \label{fig:sim_profiles}
\end{figure*}

The radial profiles of the brightness are given in Fig.~\ref{fig:sim_profiles}, the rows correspond to different angular position (as shown in the right panel of Fig.~\ref{fig:orig_vs_sim}). In the left column we show the profiles of the Level-2 image (thick curves) and the calculated diffraction (thin curves). The right column shows the profiles after diffraction subtraction. Gray dotted and solid lines show the K-corona brightness corresponding to solar maximum and minimum from \citep{Koutchmy2000}. In the plots the horizontal axes represent the distance from the IO center and are expressed in ASPIICS pixels. This distance was not converted to the distance from the solar center, because it would be different for different rows of Fig.~\ref{fig:sim_profiles}; however, the geometrical vignetting \citep{Shestov2021} is correctly calculated for synthetic K-corona profiles in each panel \citep[see ][for further details of vignetting]{2022A&A...665A.109T}. 

After diffraction is subtracted, the double-peak structure is almost completely removed from ASPIICS profiles. In particular, in the vignetting zone, the actually observed profiles correspond well to the vignetting of the model K-corona, shown by gray lines.

\begin{figure*}[!ht]
  \sidecaption
  \centering
  \includegraphics[width=12cm]{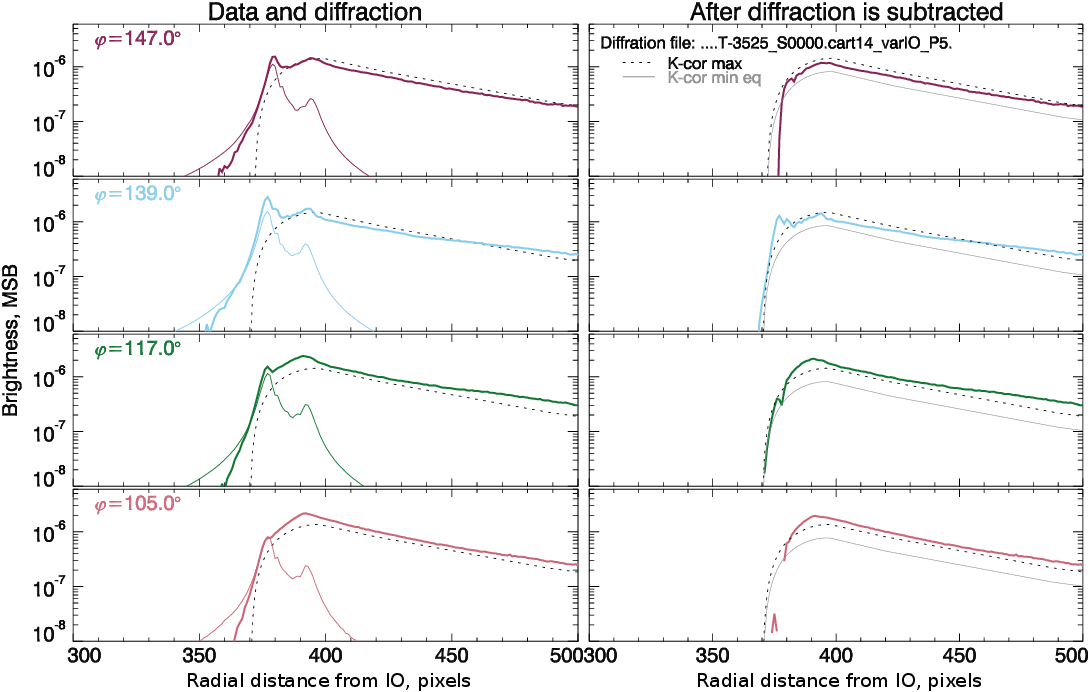}
  \caption{Similar to Fig.~\ref{fig:sim_profiles}, but for the image taken at 18:07 UTC, for which the off-pointing is +26/-35 arcsec in pitch/yaw.}
  \label{fig:orig_vs_sim_1807}
\end{figure*}

\begin{figure*}[!ht]
  \sidecaption
  \centering
  \includegraphics[width=12cm]{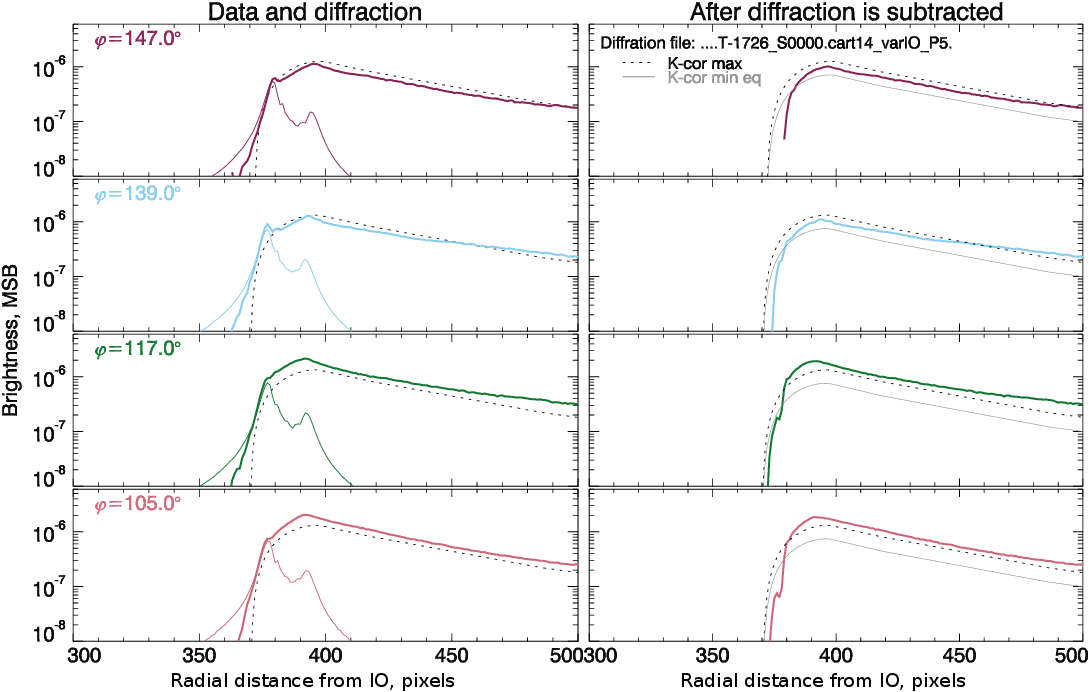}
  \caption{Similar to Fig.~\ref{fig:sim_profiles}, but for the image taken at 17:51 UTC, for which the off-pointing is +17/-26 arcsec in pitch/yaw.}
  \label{fig:orig_vs_sim_1751}
\end{figure*}

Similar radial profiles, but for the images with +35/-26 arcsec and +17/-26 arcsec off-pointings, are presented in Figs.~\ref{fig:orig_vs_sim_1807}, \ref{fig:orig_vs_sim_1751}. The overall effect of diffraction subtraction is very similar: the two-peak profile is removed, after subtraction the coronal signal shows vignetting which is very similar to the theoretically predicted.

Position of the maximum of the diffraction on $\varphi$ should depend on the combination of the pitch/yaw angles. We have analyzed angular dependence of the first and the second diffraction peaks for the files with different off-pointings and compared them with simulated diffraction profiles. In general the simulated data well represented the angular dependence, in particular position of the maximum and the presence of local variations were matching. The absolute intensity of the simulated diffraction was smaller than that of the actual files.  This partially can be explained by the contribution of the coronal signal to the registered images: at the heights of the first diffraction peak corona can reach $10^{-6}$~MSB even with the vignetting taken into account (see left panel in Fig.~\ref{fig:orig_vs_sim}). Nevertheless, variation of the brightness of the diffracted light over $\varphi$ is larger in the registered images. This also suggests that the model underestimates the diffracted signal. However, for the second diffraction peak, the difference in intensity is smaller.

Another possible cause of the discrepancy can be an imprecise knowledge of the position of the telescope aperture in the umbra-penumbra pattern. 

\section{Dependence of diffraction on other parameters}
\label{sec:others}
\subsection{Influence of the exact position of the solar center}
Since the precise position of the solar center is not known, an error of a few pixels could introduce some changes to the diffraction. Displacement of the Sun in the images corresponds to the solar shift in terms of the \SZ model. The major impact comes from the fact, that if the Sun is not co-centered with the EO, the telescope aperture is displaced from the center of the umbra-penumbra pattern. In our geometry, $\mathrm{ISD}\approx 144$~m and the pixel plate scale 2.817~arcsec, so a two-pixel displacement of the solar center in the image (5.6 arcsec) will correspond to a lateral displacement of $\sim4$~mm of the aperture, which will modify the diffraction profile. 

\begin{figure}[!h]
  \centering
  \includegraphics[width=9cm]{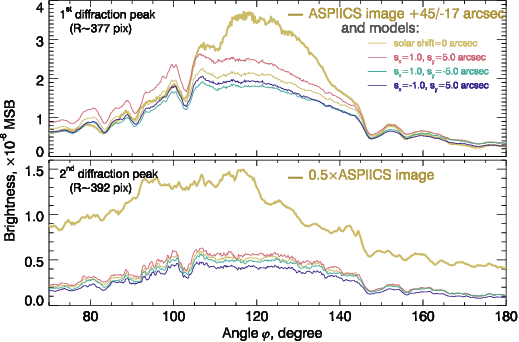}
  \caption{Influence of solar shift on diffraction model. The plots present polar dependencies of brightness in the ASPIICS image and models with various solar shift. The profiles are taken over the first (top panel) and the second (bottom panel) diffraction peaks. In the bottom panel the ASPIICS signal is $\times2$ reduced to compensate a higher contribution of the solar corona.}
  \label{fig:solar_shift}
\end{figure}

We calculated diffraction model with additional solar shift of $\pm1$ and $\pm5$~arcsec for the image with +17/-45 arcsec off-pointing. The results are presented in Fig.~\ref{fig:solar_shift}, which shows dependencies of brightness in the ASPIICS image and models over the polar angle $\varphi$. The profiles are taken over the first (top panel), and the second (bottom panel) diffraction peaks. The relative change of the most favourable to the less favourable configuration can reach 40\% in the region close to the first diffraction peak. These changes are slightly smaller already in the second peak, although it is worth to note larger relative intensity of the coronal signal.

We see that precise position of the solar center, which is accounted for in the \SZ model as solar shift, has some influence. At the current phase, we can not identify Sun's position with a higher precision; such analysis is planned for the future.
 
\subsection{Solar limb darkening}
\label{sec:limb_darkening}
In the previous analyses -- \citet{Aime2013}, \citet{Rougeot2018}, \SZ\ -- the solar limb darkening from  \citet{1993AJ....106.2096V} was used, which is a model used in astrophysics. Due to the nature of the diffraction, the light coming from the areas closer to the solar limb provides more intensity to the closest part of the diffracted ring (see Fig.~4 in~\SZ). Thus different limb darkening profiles must be giving slightly different diffraction profiles.
We compared and tried profiles from \citet{1977SoPh...51...25P} for $\lambda = 552.2$~nm expressed by 2-degree and 5-degree polynomials (P5 hereafter), from \citet{1994SoPh..153...91N} for $\lambda = 560.0$~nm (all three profiles were resembling each other despite different number of terms and $\lambda$), as well as the \citet{1993AJ....106.2096V} profile originally used in the diffraction model \citep[Eq.~2 in~][]{Rougeot2018} and the square root of it.

\begin{figure}[!h]
  \centering
  \includegraphics[width=9cm]{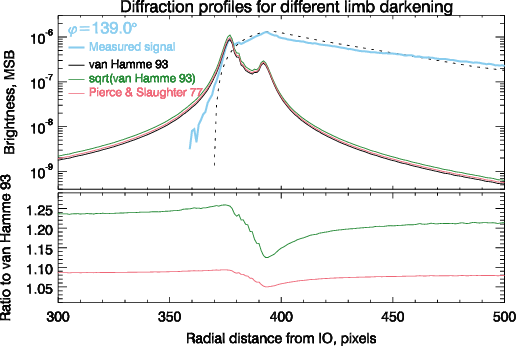}
  \caption{\textit{Top} panel: comparison of diffraction profiles for various solar limb darkening models: van Hamme 93 model, square root of it, and P5. Measured ASPIICS signal and K-corona profile from \citet{Koutchmy2000} are given by thick solid and dotted curves. \textit{Bottom} panel: the ratio of the square root to van Hamme, and P5 to van Hamme model.}
  \label{fig:limb_darkening_diff}
\end{figure}

A comparison of the diffraction for the originally used van Hamme model, the square root and the P5 model is given Fig.~\ref{fig:limb_darkening_diff}. The top panel shows the brightness measured by ASPIICS (the profile from the 2$^\mathrm{nd}$ row in Fig.~\ref{fig:orig_vs_sim}), the K-corona model during solar maximum, and corresponding diffraction profiles. The bottom panel shows the ratio of the square root to van Hamme, and P5 to the van Hamme model.

The analysis shows that introducing of a different limb darkening model may change the diffraction by $\sim10$\%. It also slightly modifies the ratio between the inner and the outer diffraction peaks. Thus, we have switched to the P5 profile for the limb darkening, whose $\lambda$ corresponds to the central wavelength of the wideband filter of ASPIICS\footnote{All the diffraction profiles reported in this paper are calculated for the wavelength $\lambda = 550$~nm.}. We note also the square root model was considered here only to demonstration purposes and is not used elsewhere. 

\subsection{Influence of the inter-satellite distance}
Changing of the ISD has also a slight influence on the diffraction. While the on-board systems control and measure the variation of the ISD with millimeter precision, we evaluate how much a larger variation of the ISD would affect the diffraction pattern. 
We have calculated diffraction for the same off-pointing +17/-45 arcsec but using $z_0=144.23$~m and $z_0=145.03$~m in addition to the baseline value of $z_0=144.63$~m. These $\sim40$~cm changes in ISD resulted in almost constant (along radius) increase or decrease of the diffracted signal by 10\% with respect to the baseline value. With larger ISD the diffracted signal was smaller, while in was larger with the smaller ISD. This looks counter-intuitive: with increase of ISD the apparent angular size of the EO is reduced, which should increase the amount of the diffracted light coming to the telescope. However, with larger ISD the size of the bright diffracted ring in the primary focus is reduced, so the IO blocks the diffracted light with a higher efficiency. 

\section{Diffraction in regular ASPIICS images}
\label{sec:regular}
During nominal observations, the ASPIICS images are registered with a much better co-alignment: ASPIICS points within a few arcsec to the EO center, and the solar center coincides with the EO center within a few arcsec. The combination of these two factors results in the tilt/shift angles  of the order of 5--10 arcsec. For such off-pointings, the diffracted light has a more symmetrical shape and the brightness is smaller than those presented in Sect.~\ref{sec:refined}, since the IO blocks the  diffraction ring with a higher efficiency. 

\begin{figure}[ht]
  \centering
  \includegraphics[width=9cm]{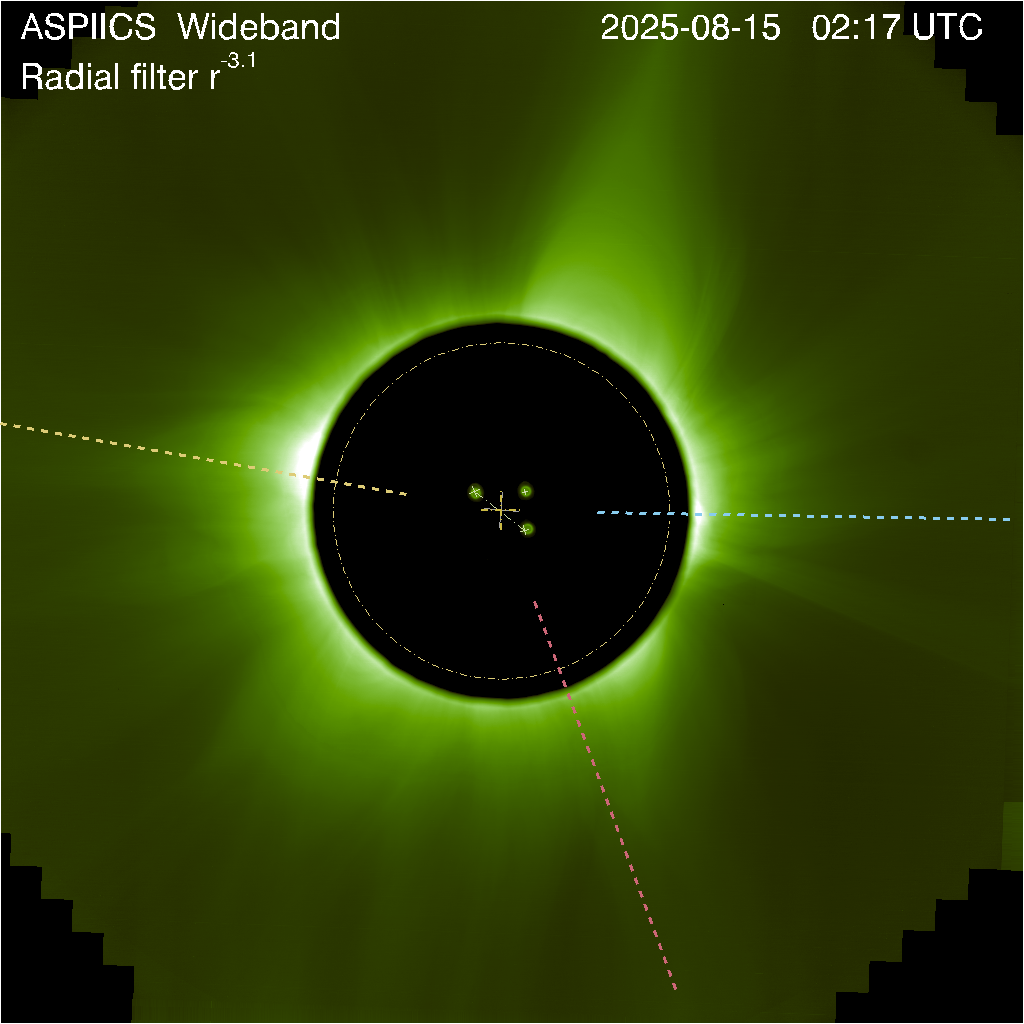}
  \caption{Wideband ASPIICS image from orbit 309 registered around 02:17 UTC on 15 Aug 2025. This is a Level-3 image, composed from three individual images with $t_\mathrm{exp}=0.1$, 1.0, and 10.0~s. A radial filter with factor $r^{-3.1}$ is applied to the image to enhance visibility. The image is Sun-centered and de-rotated such that the solar north points to the top of the figure. Dashed lines denote positions of profiles shown in Fig.~\ref{fig:orbit309_profile}.}
  \label{fig:orbit309_image}
\end{figure}

\begin{figure}[ht]
  \centering
  \includegraphics[width=9cm]{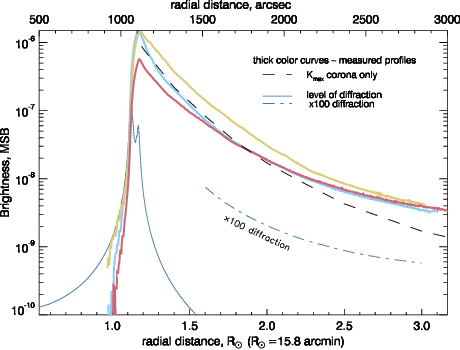}
  \caption{Comparison of the coronal brightness and diffracted light. Thick colored curves show the radial profiles of the coronal brightness along the lines in Fig.~\ref{fig:orbit309_image}. The solid blue line corresponds to the diffracted light calculated by our refined model, and the dash-dotted blue line shows the same diffracted light intensity multiplied by a factor 100. The black dashed line denotes the K-corona brightness during solar maximum \citep{Koutchmy2000}.}
  \label{fig:orbit309_profile}
\end{figure}

An example of a typical Level-3 ASPIICS image is given in Fig.~\ref{fig:orbit309_image}: a wideband image from orbit 309 on 15 August 2025, registered at 02:17 UTC. This Level-3 image was assembled from three Level-2 images, taken with $t_\mathrm{exp}=0.1$, 1.0, and 10.0~s. A radial filter with factor $r^{-3.1}$ was applied to the image to enhance visibility of strucures. The image is Sun-centered and de-rotated, such that the solar north points to the top. The dashed lines denote position of the radial profiles: the blue line crosses open fields, the yellow line crosses a streamer, and the red line crosses a quiet Sun region. The radial profiles are shown in Fig.~\ref{fig:orbit309_profile} with thick colored curves (the colors correspond to the lines in Fig.~\ref{fig:orbit309_image}). Since the center of the Sun coincides with a reasonable accuracy with the IO center, the radial distances are expressed in $R_\sun$. The dash-dotted blue line starting at $\sim1.6R_\sun$ shows the intensity of the diffracted light in the outer field of view multiplied by a factor 100. The simulated diffraction is significantly dimmer than the corona. Even in the most of the vignetting zone, the intensity of the diffracted light is an order of magnitude lower than the coronal brightness. Thus the double-peaked structure of the diffraction ring is not noticeable neither in the radial profiles, nor in the registered image.

\section{Discussion and conclusions}
\label{sec:conclusions}
We presented several ASPIICS observations taken with off-pointings, which quantitatively confirm all the properties of the diffracted light predicted by the analytical-numerical models summarized in Sect.~\ref{sec:model}. For large off-pointings, when the IO does not block the diffraction ring produced by the EO, the image of this diffraction ring is directly registered on the detector. The radial profiles of the observed and modeled diffraction ring match well. The peak intensity given by the model is $\sim30\%$ higher, which is probably due to superior performance of the toroidal shape of the EO \citep{Landini2011}. However, the numerical modeling of such three-dimensional occulters is quite challenging  \citep{Aime2020}.

This 30\% discrepancy is a reasonably good coincidence given all the approximations of the model. It serves also as a validation of the  radiometric calibration of ASPIICS. The brightness profiles presented in this paper are expressed in units of MSB, and these are the natural units for the diffraction model. The ASPIICS images were initially expressed in DNs, and during the Level-2 processing (which included also other transformations) were converted to MSB using radiometric calibration factors. These calibration factors were derived during the on-ground calibrations of ASPIICS from the combination of the measured radiometric sensitivity expressed in W cm$^{-2}$ sr$^{-1}$ DN$^{-1}$, and the convolution of the known spectral profiles of the ASPIICS passbands with the solar spectrum \citep{Zhukov2025}.  Thus the correspondence of the observed and modeled intensities of the diffracted light validates also the radiometric sensitivity values and the whole approach for calibration. 

We refined the numerical model such that it better reproduces the geometrical properties seen in the images registered with moderate ASPIICS off-pointings. In particular, the double-peak diffraction profile and  the position of its maximum depending on the off-pointings are well reproduced.  The need for these changes might be related to the errors introduced by simplified geometry of the diffraction model (such as infinitely thin lenses). The model was overestimating the intensities for the diffracted light at large off-pointings, and underestimating them for smaller off-pointings. This  suggests that an error in the radiometric sensitivity is an unlikely source of the discrepancy. A part of the discrepancy might be accounted for by the position of the solar center that is not precise enough for the model (such analysis is planned for future). Other unaccounted physical effects that may contribute to the discrepancy include the finite PSF of the PO and thus a less efficient rejection by the IO, or the finite transparency of the IO at its edge. 

It is worth to mention that the diffraction model used a monochromatic approach with $\lambda=550$~nm. At the same time, the model predicts that that the intensity of the diffracted light should slightly depend on wavelength, with the intensity in the plane A inversely proportional to $\lambda$. With the ASPIICS passbands centered at $\lambda=551$~nm (wideband and polarized channels), 530.3~nm (Fe XIV channel), and 587.6~nm (He I channel), the wavelength dependence could contribute additional 5--7\% of discrepancy. There are other mechanisms that lead to the diffracted light dependence on $\lambda$, which, however, we do not expect to be significant. As discussed in \SZ, it is the rejection by the IO that determines the suppression of the diffracted light, and this rejection depends on the size of the bright diffracted ring formed by the PO, and its relative position with respect to the IO, which both should not depend on wavelength. The lenses may suffer from small chromatic aberrations, so the rejection efficiency may be smaller. These factors will be investigated during further analysis of ASPIICS data. 

One of the main results of the diffraction model is that the intensity of the diffracted light is significantly reduced when ASPIICS is well co-centered and co-aligned with the Sun, the EO and the umbra-penumbra pattern. During nominal observations the co-alignment of ASPIICS, the EO, and the solar center is very good, so the intensity of the diffracted light is at least 2 orders of magnitude below the coronal signal. In the outer part of the vignetting zone, the contribution of the diffraction can be up to 10\%.

\section{Data availibility}
The ASPIICS data is freely available at the website of the Proba-3/ASPIICS Science Center \url{https://www.sidc.be/proba-3/aspiics-data}

The CPU-based diffraction code is available at: \url{https://gitlab.com/sshestov/aspiics_diffraction}

\begin{acknowledgements}
      {The ASPIICS data are courtesy of the Proba-3/ASPIICS consortium. Proba-3 is a technology demonstration mission of the European Space Agency (ESA) and a Mission of Opportunity in the ESA Science Programme. The ASPIICS project is developed under the ESA’s General Support Technology Programme (GSTP) and the ESA’s PRODEX Programme thanks to the contributions of Belgium, Poland, Romania, Italy, Ireland, Greece, and the Czech Republic. The ROB team thanks the Belgian Federal Science Policy Office (BELSPO) for the provision of financial support in the framework of the PRODEX Programme of ESA under contract numbers 4000117262, 4000136424, 4000145189, and 4000147286. S.S. acknowledges the Belgian FED-tWIN program, useful discussions within the ISSI International Team project 23-572 on Models and Observations of the Middle Corona, funded by the International Space Science Institute (ISSI) in Bern; he is also grateful to Abderraouf Yamani for his support. S.G. acknowledges the support from grant 25-18282S of the Czech Science Foundation (GA\v CR). P.L. acknowledges financial support from Centre National d'Etudes Spatiales. The reviewer is thanked for useful comments on this manuscript.      }
\end{acknowledgements}

\bibliographystyle{aa}
\bibliography{aspiics_diffraction}

\begin{appendix}

\section{Details of mathematical approach}
    \label{sec:Ap_method}
    In our analysis we follow the method of \citet{Aime2013} and \citet{RR17}. Firstly we consider propagation of a wave
    $\Psi$ through an optical system \citep{goodman2005introduction}. After propagation through some aperture $A$, at distance $z$ the wave amplitude $\Psi$ is calculated
    as:
    \begin{multline}
      \Psi_z (x,y) = \frac{\exp \left( \frac{i\pi (x^2+y^2)}{\lambda z} \right)}{i\lambda z} \mathcal F  
	    \left\{ A \cdot \Psi(\xi,\eta) \cdot \right. \\
        \left. \cdot \exp \left(\frac{i \pi (\xi^2+\eta^2)}{\lambda z} \right) \right\},
    \end{multline}
    where $(x,y)$ --- coordinates in the observer plane, $\mathcal F$ --- Fourier transform, $(\xi,\eta)$ --- coordinates in the aperture plane, 
    $\lambda$ --- wavelength, $A = A(\xi,\eta)$ --- aperture function that denotes transparency at point $(\xi, \eta)$. 

    If one places a convergent lens with the focal distance $f$ immediately after the aperture, an additional term $\exp \left( -
    \frac{i \pi (\xi^2+\eta^2)}{\lambda f} \right)$ must be added inside the $\mathcal F$ argument. In the case $z$ and $f$ coincide we obtain a
    known expression for a focal plane of the lens:
    \begin{equation}
      \Psi_z (x,y) = \frac{\exp \left( \frac{i\pi (x^2+y^2)}{\lambda z} \right)}{i\lambda z} \mathcal F \left\{ A \cdot \Psi(\xi,\eta) \right\}.
    \end{equation}

    With this we write an expression for a wave propagating from the aperture to the detector:
    \begin{multline}
      \Psi_D (x,y) = \frac{\exp \left( \frac{i \pi (x^2+y^2)}{\lambda l} \right)}{i \lambda l} 
      \cdot \mathcal F \left[ A_C \cdot \Psi_{C} \cdot \right. \\
      \left. \cdot \exp \left( \frac{i \pi r^2}{\lambda l}  \right) \right],
    \label{PsiD}
    \end{multline}
    where $l$ is distance between O3 and $D$, $ \Psi_C = \mathcal F \left\{ A_{O^\prime} \cdot \Psi_{O^\prime} \right\}$, and $\Psi_{O^\prime} = \mathcal F \left\{ A_A \cdot \Psi_A \cdot \exp
      \left( - \frac{i \pi r^2}{\lambda z_0}  \right)  \right\}$ (for the latter we used the relation between $z_0$ and $z_1$).

    We note that arguments to each Fourier transform are taken at a corresponding plane and the
    coordinates and linear scales are different in each case. To perform numerical modeling of the Fourier propagation,
    we substitute the 2D functions $\Psi_A$, $A_A$, $\Psi_{O'}$, $A_{O'}$, $\exp \left( \frac{i \pi r^2}{\lambda z_1}
    \right)$, \emph{etc.} by corresponding 2D arrays.

    Now we consider diffraction of a plane-parallel wave on a infinitely thin circular occulter. The problem has been considered by \citet{Aime2013}, with the result that the amplitude can be expressed by the amplitude of initially co-axial wave $\Psi_{A00}$ (see his Eq.~5):
    \begin{equation}
	\Psi_{A \alpha \beta} (x,y) = 
	  T_{\alpha \beta}(x,y) \cdot \Gamma_{\alpha \beta}(x,y) \cdot \Psi_{A00} (x+\alpha z_0, y+\beta z_0), 
	\label{PsiA}
    \end{equation}
    where $\Psi_{A00} (\xi,\eta)$ --- amplitude of initially co-axial wave, and 
    \begin{gather}
      T_{\alpha \beta} (x,y) = \exp \left( -2 \pi i \frac{\alpha x + \beta y}{\lambda} \right) \qquad \mathrm{(Tilt)} \label{Tab} \\
      \Gamma_{\alpha \beta} (x,y) = \exp \left( -\pi i \frac{ (\alpha^2 + \beta^2) z_0}{\lambda} \right) \qquad \mathrm{(Offset)}, \label{Gab}
    \end{gather}
    $\Psi_{A00} (\xi,\eta)$ is calculated using Fourier-Hankel transform:
    \begin{equation}
       \Psi_{A00} (\xi,\eta) = 1- \frac{\varphi_{z0}(r)}{i \lambda z_0} \int_0^R 2 \pi \rho \exp
       \left( i \pi \frac{\rho^2}{\lambda z_0} \right) J_0 \left( 2\pi \frac{r \rho}{\lambda z_0} \right) \, d\rho,
       \label{PsiA00}
    \end{equation}
    where $r=\sqrt{\xi^2 + \eta^2}$ --- radial coordinate in the $A$ plane, $\varphi_{z0} = \exp (i\pi r^2/ \lambda
    z_0)$, and $J_0(r)$ is the Bessel function of the first kind. The function $\Psi_{A00}$ is circularly symmetrical
    and is calculated initially as a 1D array.

\end{appendix}

\end{document}